\documentclass[fleqn,10pt]{wlscirep}
\usepackage[utf8]{inputenc}
\usepackage[T1]{fontenc}
\usepackage{color}

\def\P{{\mathbb P}}

\title{Morphological organization of point-to-point transport in complex networks}

\author[1,+]{Min-Yeong Kang}
\author[2,3,4,+]{Geoffroy Berthelot}
\author[1,5,+]{Liubov Tupikina}
\author[6,7]{Christos Nicolaides}
\author[2]{Jean-Francois Colonna}
\author[1]{Bernard Sapoval}
\author[1,*]{Denis S. Grebenkov}

\affil[1]{Laboratoire de Physique de la Mati\`{e}re Condens\'{e}e (UMR 7643),
CNRS -- Ecole Polytechnique, IP Paris, 91128 Palaiseau, France}

\affil[2]{Centre de Math\'ematiques Appliqu\'ees, CNRS -- Ecole Polytechnique, IP Paris,
91128 Palaiseau, France}

\affil[3]{Research Laboratory for Interdisciplinary Studies (RELAIS), 75012 Paris, France}

\affil[4]{Institut National du Sport, de l'Expertise et de la Performance (INSEP), 75012 Paris, France}

\affil[5]{The Center for Research and Interdisciplinarity (CRI), University Paris Descartes, INSERM, France}

\affil[6]{Department of Business and Public Administration, University of Cyprus, 1 Panepistimiou Av. 2109 Aglantzia, Nicosia, Cyprus}

\affil[7]{MIT Sloan School of Management, 100 Main Street, Cambridge, MA 02142, USA}

\affil[*]{denis.grebenkov@polytechnique.edu}

\affil[+]{these authors contributed equally to this work}


\date{\today}

\begin{abstract}
We investigate the structural organization of the point-to-point electric, diffusive or hydraulic transport in complex scale-free networks. The random choice of two nodes, a source and a drain, to which a potential difference is applied, selects two tree-like structures, one emerging from the source and the other converging to the drain.  These trees merge into a large cluster of the remaining nodes that is found to be quasi-equipotential and thus presents almost no resistance to transport.  Such a global ``tree-cluster-tree'' structure is universal and leads to a power law decay of the currents distribution.  Its exponent, $-2$, is determined by the multiplicative decrease of currents at successive branching points of a tree and is found to be independent of the network connectivity degree and resistance distribution.
\end{abstract}

\begin{document}

\flushbottom
\maketitle
%
%
\thispagestyle{empty}


\section*{Introduction}

Transport processes in complex networks play a crucial role in our
lives, with examples ranging from transportation networks (e.g.,
airflight/train connections, international highways, and city
transport), electricity distribution networks, microelectronic devices
to news propagation and spreading of behaviors in social networks
\cite{Barthelemy2000,Havlin,Pastor,Caldarelli,Condamin2007,Gallos2007a,Corson2010,Li2010,Mulken2011,Nicolaides2012,Aral2017}.
In engineering and biological applications, transport is often
electric, hydrodynamic, or diffusive
\cite{Hernandez,Mauroy2004,Grebenkov2005,Hu2013}.  During the past two
decades, a particular attention has been paid to random networks which
can capture major structural properties of complex systems
\cite{Albert2002,Goh2002,Satoras,Gallos2007,Radicci2016}.  In particular,
scale-free networks, in which the degree distribution follows a power
law $P(k) \propto k^{-\gamma}$ with an exponent $\gamma$, can model
various systems with scaling properties such as citation patterns in
science \cite{Barabasi1999}, internet \cite{Dorogov2002}, e-mail
connections \cite{Seyed2006}, to name but a few.  Scale-free networks
display a variety of interesting transport features found in nature
\cite{Bollt2005,Dorogov2002}.  In particular, the global
point-to-point flux $\Phi$ in a random scale-free resistor network
between two arbitrarily chosen points was shown to obey a distribution
with a power law tail: $P(\Phi) \propto \Phi^{-(2\gamma - 1)}$
\cite{Lopez2004}.  The degree exponent $\gamma$ was linked to the
scaling of the flux in the line-to-line transport in a resistor
network with multiple sources and drains \cite{Nicolaides2010}.

In a typical setting, a particle, an animal, a disease, a virus, a
toxin, a signal or a rumor is released at one location and then
spreads over the network.  To get a theoretical insight onto this
phenomenon, a first step is to consider the
\emph{point-to-point transport} between two arbitrary nodes, treated
as a source and a drain.  In this paper, we investigate the electric
transport in a class of scale-free resistor networks, obeying
Kirchhoff's laws or, equivalently, a Markov chain random walk on
weighted graphs.  This study reveals the structural organization of
the nodes potentials and currents in the links.  It is found that the
density of currents, $p(\phi)$, decays at large $\phi$ as a power law
with the universal exponent $-2$.  This behavior is attributed to the
arborescent structure of links that drive the currents from the source
to the drain through a large {\it quasi-equipotential (QEP) cluster}
of nodes.  In other words, the transport between two selected nodes of
a network is governed by two random trees, for which the exponent $-2$
can be justified by theoretical arguments.

\section*{Results}

A random scale-free network is constructed on an $N\times N$ square
lattice.  Its links are generated by using an uncorrelated
configuration model \cite{Catanzaro2005} with a given degree exponent
$\gamma$.  To get more insight onto the universal character of the
point-to-point transport, we consider the resistance $r_{i,j}$ of each
link as a function of the Euclidean distance $d_{i,j}$ between the
nodes $i$ and $j$: $r_{i,j} = d_{i,j}^{\beta}$, with an exponent
$\beta$.  In an electric or hydraulic circuit, the resistance of a
wire or a tube is proportional to its length, and $\beta = 1$.  In
turn, most former studies on transport in resistor networks supposed
constant link resistance, i.e., $\beta = 0$ (see, e.g.,
\cite{Lopez2004}).  In each random realization of the resistor network
with prescribed exponents $\gamma$ and $\beta$, we select randomly a
source node and a drain node, at which the potential is fixed to be
$1$ and $0$, respectively.  The system of linear Kirchhoff's equations
\cite{Redner2001} for the potential on other nodes is solved
numerically using a custom routine in Matlab (see Methods section).
Then the distributions of nodes potentials and currents in the links
are obtained.  While we keep using the terminology of electric
circuits, the results are valid for diffusive and hydraulic transport
as well.

\subsection*{Universal power-law distribution of currents}

\begin{figure}
\centering
\includegraphics[width=170mm]{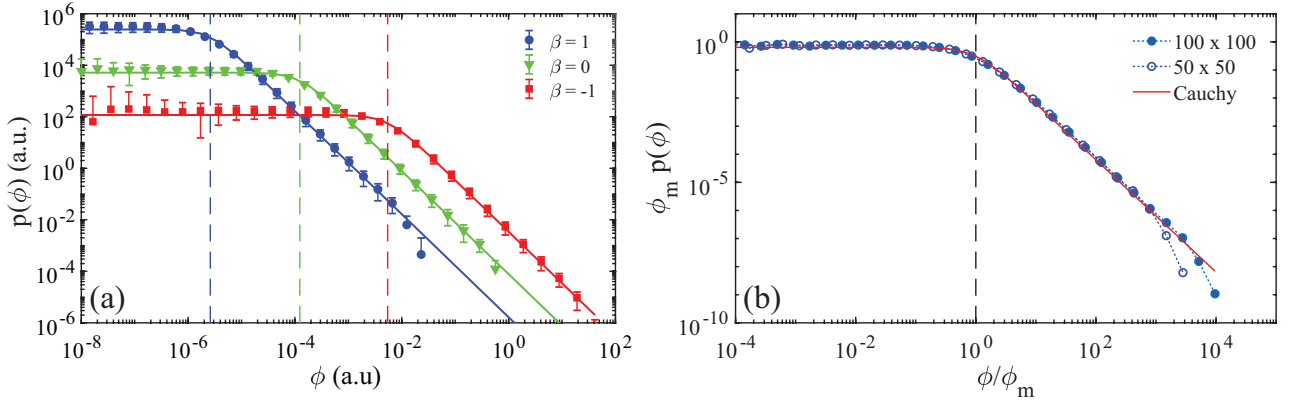}
\caption{
{\bf (a)} The density of currents in links of a scale-free resistor
network on a lattice $100 \times 100$, with $\gamma = 3$ and three
values of $\beta$.  The empirical probability densities were first
estimated from a set of currents in each random realization of the
network and then averaged over 300 realizations.  Symbols show the
average values while errorbars indicate the average plus or minus the
standard deviation.  The bottom part of the errorbar is missing
whenever the average minus the standard deviation is negative.  The
one-sided Cauchy density in Eq. (\ref{eq:Cauchy}) with the median
currents from Table \ref{tab:parameters} is shown by solid lines.
Dashed vertical lines indicate the median currents $\phi_m$ for each
$\beta$.  Similar results were obtained for other values of $\gamma$
(not shown). {\bf (b)} The empirical density of currents rescaled by
the median current $\phi_m$ for a scale-free resistor network with
$\gamma = 3$, $\beta = 1$, $N = 100$ (full circles) and $N = 50$
(empty circles), as compared to the one-sided Cauchy density (solid
line).  Here, the empirical probability densities were obtained by
merging local currents from 300 random realizations of the network to
improve the statistics (errorbars are not available here). }
\label{fig_beta_gamma}
\end{figure}

The first striking result is that the distribution of currents in a
random scale-free resistor network follows a universal power law
distribution, that is independent of the degree distribution exponent
$\gamma$ and the resistance-distance exponent $\beta$.
Figure~\ref{fig_beta_gamma}(a) shows that the density of currents is
constant at small currents and then decays as a power law with the
exponent $-2$.  This probability density admits an excellent fit by
the one-sided Cauchy distribution (for $\phi \geq 0$),
\begin{equation}  \label{eq:Cauchy}
p(\phi) = \frac{2}{\pi} \, \frac{\phi_m}{\phi_m^2 + \phi^2} \,,
\end{equation}
where $\phi_m$ is the median current.  Interestingly, the mean and the
variance of this distribution are infinite for an infinite network,
which is a reminiscent property of scale-free systems.
The universal power law decay can potentially be related to formerly
studied betweenness measures in scale-free networks
\cite{Freeman1977,Barthelemy2004,Newman2005}.
While the distribution is universal, the median current $\phi_m$
depends on the properties of the network (Table \ref{tab:parameters}).
The median current $\phi_m$ has a relatively weak dependence on
$\gamma$, as compared to substantial changes of the properties of the
scale-free network for various $\gamma$.  The much stronger dependence
on $\beta$ results from a significant change of the length-dependent
resistances: when $\beta = 1$, resistances are proportional to
inter-node distances and are thus large (yielding small currents); in
turn, for $\beta = -1$, the resistances are much smaller while the
currents are larger.  The deviations from the power law decay $p(\phi)
\propto \phi^{-2}$ at large currents can be attributed to a finite
size of the system.  In fact, for any finite-size resistor network,
the inter-node distances are bounded from above and below, and thus
there exists a minimal resistance and a maximal current.  As a
consequence, the distribution of currents has a finite support bounded
by a maximal cut-off.  The cut-off value depends on the exponent
$\beta$ and the size $N$ of the system, and increases as $N\to\infty$.
This is clearly seen on Fig.~\ref{fig_beta_gamma}(b), which shows the
density of currents rescaled by the median current $\phi_m$ for $N =
50$ and $N = 100$.  The rescaling is needed to make closer the
densities for two cases because the median current depends on the
system size.  One can see that the empirical density for the larger
system remains close to the theoretical curve up to larger currents,
i.e., the cut-off current is larger.

\begin{table}
\begin{center}
\begin{tabular}{|c|c|c|c|}  \hline
$\gamma ~ \backslash ~ \beta$ & $1$ & $0$ & $-1$  \\  \hline
1.5 & 0.56 & 0.27 & 1.26 \\
2.0 & 1.21 & 0.58 & 2.48 \\
2.5 & 2.02 & 0.92 & 4.14 \\
3.0 & 2.40 & 1.16 & 5.00 \\
4.0 & 2.99 & 1.52 & 5.77 \\
5.0 & 3.52 & 1.73 & 6.58 \\
6.0 & 3.74 & 1.91 & 6.95 \\  \hline
    & $\times 10^{-6}$ & $\times 10^{-4}$ & $\times 10^{-3}$  \\ \hline
\end{tabular}
\end{center}
\caption{
The median current $\phi_m$ for different exponents $\gamma$ and
$\beta$.  For each value of $\gamma$ and $\beta$, the empirical
distribution was obtained by merging currents from 300 random
realizations of scale-free networks on a lattice $100\times 100$.}
\label{tab:parameters}
\end{table}

\subsection*{Morphological organization of the point-to-point transport}

\begin{figure*}
\begin{center}
\includegraphics[width=170mm]{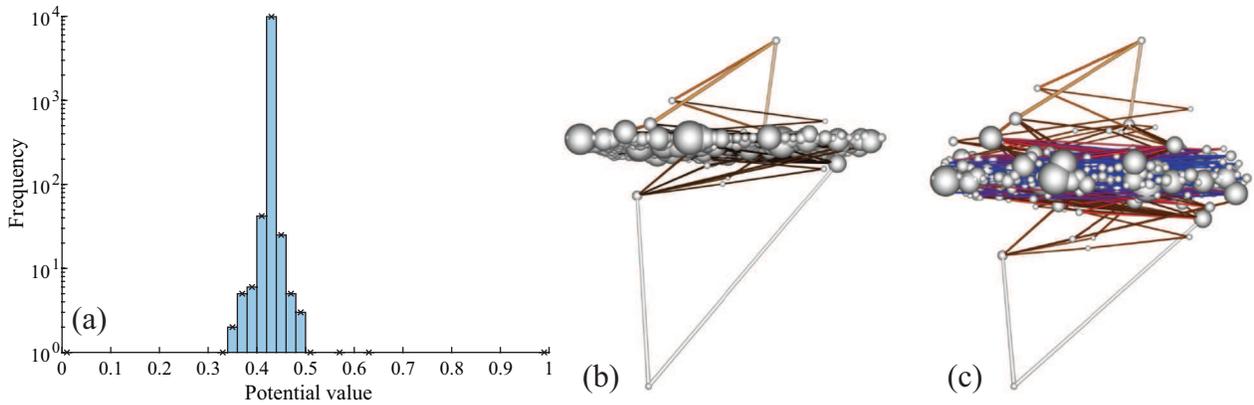}
\end{center}
\caption{
{\bf (a)} Histogram of the nodes potentials (lattice $100\times100$,
$\beta=1$, and $\gamma=2.5$).  One observes that the overwhelming
majority of nodes have almost identical potential.  {\bf (b,c)}
Visualization of a point-to-point transport on a scale-free network
(with a lattice $40\times 40$, $\gamma=2.5$ and $\beta=1$; high
resolution images are available online).  Each node of the network is
shown by a ball whose radius is proportional to the square root of its
connectivity.  The planar coordinates of the balls are the positions
of the corresponding nodes on the square lattice, whereas the height
$Z$ is related to the potential at the node, either by a linear
relation $Z = V$ {\bf (b)} or by a nonlinear relation $Z
\propto {\rm sign}(V-V_m) |V - V_m|^{1/2}$ {\bf (c)}, where $V_m$ is
the mode of the potential histogram, and rescaling is used to ensure
that the bottom black ball at $Z=0$ is the drain at fixed potential
$0$ and the top white ball at $Z=1$ is the source at fixed potential
$1$.  Such a nonlinear relation helps to dilate the QEP cluster to
visualize its structure.  Each link brightness is proportional to the
magnitude of its current (in addition, blue colors are used for very
small currents in panel {\bf (c)}).  The QEP cluster is qualitatively
identified as a large ensemble of nodes almost at the same potential.}
\label{fig_potent}
\end{figure*}

To get a deeper insight onto the above transport properties, we study
the distribution of the nodes potentials.  By construction, the
potential varies between $0$ (the drain) to $1$ (the source).  Figure
\ref{fig_potent}(a) shows an example of the potential distribution for
one realization of the scale-free network with $\gamma = 2.5$ and
$\beta = 1$.  The most striking feature is the very peaked
distribution of potentials around a particular value (here, $0.43$).
In other words, the overwhelming majority of nodes share very close
potentials and form thus a quasi-equipotential cluster, which works as
an almost perfect conductor.  This analysis gives a more precise
description of the so-called transport backbone \cite{Lopez2004}.
While the particular value of the potential of the QEP cluster is
specific to the network realization, the very strong concentration
around this particular value is universal, being observed for other
values of $\gamma$ and $\beta$.

The structural organization of the point-to-point transport is
visualized in Fig.~\ref{fig_potent}(b,c).  Here the nodes of a
scale-free network are elevated in the vertical direction according to
their potential.  One can distinguish the QEP cluster (the region in
the middle), and two arborescent structures, one rooted in the source
at the top, and the other rooted in the drain at the bottom.  The
``terminal branches'' of both trees are connected to the QEP cluster
and can thus be considered as grounded to the potential of the QEP
cluster.  This structural organization allows one to treat the
point-to-point transport on networks as a series connection of two
random trees.  As the electric properties of a tree can be determined
via exact recursive computations (see below), this is a tremendous
simplification of the original problem.  We will show how this
discovery brings new conceptual understanding of the point-to-point
transport on networks and a rational for the observed universality in
the currents distribution.
The almost flat region of the density of currents in
Fig.~\ref{fig_beta_gamma} can be attributed to the small currents in
links between the numerous nodes and loops of the QEP cluster.  In
turn, the larger currents flow in two trees (see
Fig.~\ref{fig_potent}(b,c)) and produce a power law decay of the
currents distribution.  This is typical for a tree, in which the
largest current in the trunk is progressively divided in a large
number of smaller currents following the successive branching points.
As both trees are connected to the QEP cluster, they can be treated
independently.

\subsection*{Recursive computation for a tree}

\begin{figure}
\begin{center}
\includegraphics[width=85mm]{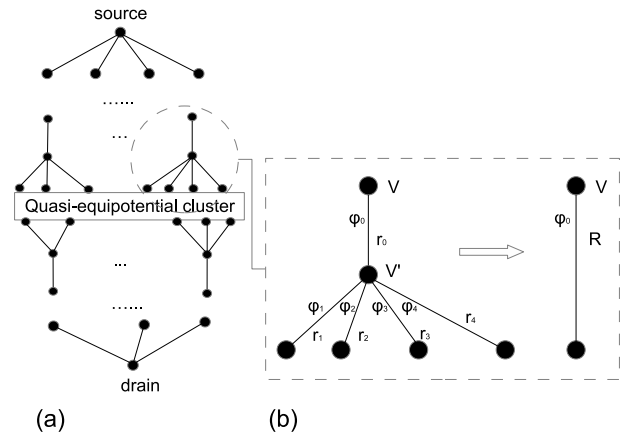}
\end{center}
\caption{
{\bf (a)} Schematic representation of the two-sided arborescent
structure created by applying potentials to two points: one tree is
rooted in the drain at the bottom and includes nodes with distinct
potentials, and the second random tree is rooted in the source on the
top and also includes nodes with distinct potentials.  The ``terminal
branches'' of both trees are connected to the QEP cluster.  If the
latter is substituted by an equipotential one, the currents can be
computed exactly by a recursive procedure.  {\bf (b)}.  Each step
of the recursive procedure consists in substituting a group of nodes
with resistances $r_0, r_1, \ldots, r_m$ by an effective link with the
overall resistance $R$.  Repeating this step, one evaluates the total
resistance and then all intermediate potentials and currents.  }
\label{fig:scheme}
\end{figure}

For an arbitrary resistors tree, in which all terminal
branches are grounded (set to a potential $V_0$) and a given potential
$V_1$ is applied to the root, the currents and potentials can be
computed via an exact recursive procedure.  In fact, for any terminal
branch, one can identify its mother node and all sister nodes
(Fig. \ref{fig:scheme}).  If $r_1, \ldots, r_m$ denote the resistances
of the links from these terminal branches to their mother node, and
$r_0$ is the resistance of the link from the mother node to its mother
node, then the overall resistance of this group is simply $R = r_0 +
1/(1/r_1 + \ldots + 1/r_m)$.  As a consequence, this group of links
can be replaced by a single effective link with the resistance $R$.
Repeating this replacement procedure from the most distant terminal
branches and progressing toward the root, we can compute the total
resistance of the tree $R_{\rm tree}$, from which we get the local
current $(V_1-V_0)/R_{\rm tree}$ at an effective link representing the
whole tree, and the potential drops at the daughter nodes.
Considering now each daughter node as the root of the corresponding
subtree, one can repeat the computation.  Descending recursively from
the root to the most distant terminal branches, one evaluates all
currents and potentials.

\begin{figure}
\centering
\includegraphics[width=136mm]{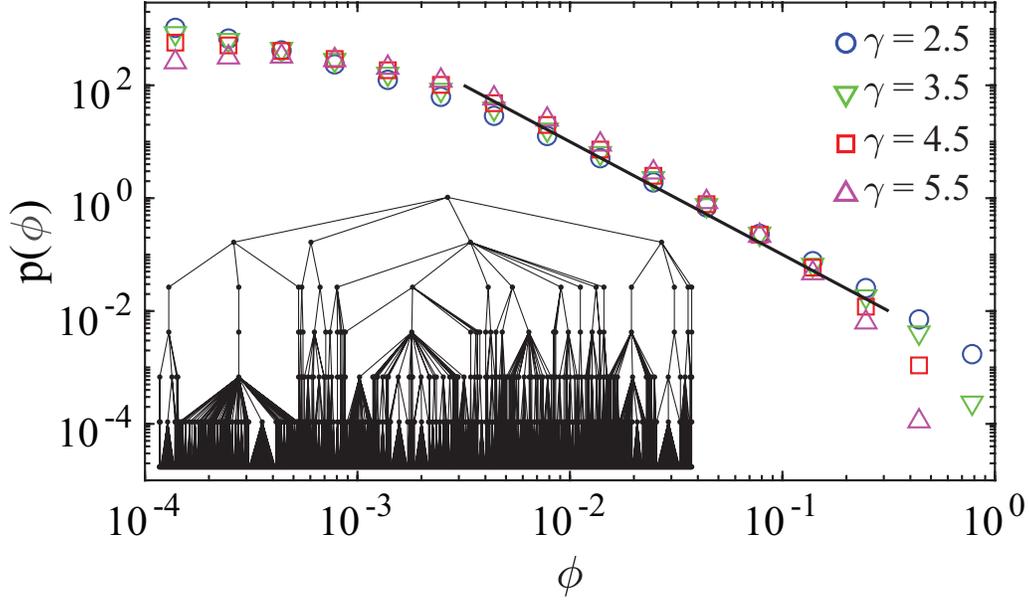}
\caption{
The density of currents in random trees built by choosing randomly the
node degrees from the degree distribution for several $\gamma$ values
(within a $100 \times 100$ lattice).  To keep the total number of
nodes in the trees in a range between $1000$ and $3000$, the number of
generations $G$ (i.e., the distance from the root to any terminal
node), was assigned differently for different $\gamma$: $G = 7$
($\gamma = 2.5$), $G = 12$ ($\gamma = 3.5$), $G = 19$ ($\gamma =
4.5$), and $G = 27$ ($\gamma = 5.5$).  Currents were computed by
fixing the potential to $1$ at the source and $0$ at the the last
generation nodes.  All branches have unit resistance (i.e., $\beta =
0$).  The density is obtained from a histogram of currents from 300
realizations.  Irrespective of $\gamma$ values, the density of
currents decays as a power law with an exponent $-2$, indicated with
the straight line (the power law is terminated by a cut-off due to a
finite size of the tree).  The inset shows an example of a random tree
with $\gamma = 2.5$. }
\label{fig:tree}
\end{figure}

At any level, once the potential $V$ at the mother of the mother node
is evaluated for a given group, we can compute the current $\phi_0 =
V/R$, the potential
\begin{equation*}
V' = V - \phi_0 r_0 = V (1 - r_0/R) = \frac{V}{1 + \frac{r_0}{r_1} + \ldots + \frac{r_0}{r_m}} \,,
\end{equation*}
and the currents $\phi_i = V'/r_i$ at each link.  If the resistances
$r_i$ are comparable to each other, then the currents $\phi_i$ in the
links to daughter nodes are approximately $m+1$ times smaller than the
current $\phi_0$, while the number of such currents is $m$ times
larger.  Most importantly, due to the arborescent character of the
structure, this reduction of currents is repeated multiplicatively at
all branching levels.  Therefore, the probability that the local
current $\hat{\phi}$ at a randomly chosen link exceeds a prescribed
value $\phi$, is dominated by the relative fraction of links with the
current of order $\phi$, which is inversely proportional to $\phi$.
In other words, $\P\{ \hat{\phi} \geq \phi\}
\propto 1/\phi$, from which the scaling $\phi^{-2}$ of the probability
density $p(\phi)$ follows immediately.  These qualitative arguments
are confirmed by a more rigorous computation for some classes of
random trees (see Methods section) and by numerical simulations for
various random trees (see Fig. \ref{fig:tree}).

\section*{Discussion}

The emergence of a ``working tree'' in a complex network is not a
surprise since the currents follow the links that were built
independently of the choice of source and drain.  Starting for the
source node, there exits a finite number of links leading to other
nodes.  At this stage there are no loops.  Then starting from these
new nodes, one finds new links but the probability of creating loops
is very small because the great majority of the support sites are free
for new links.  This process builds progressively a tree structure up
to the situation where loops are necessarily created when the number
of tree sites is comparable to total number of support sites.  This is
why the numerical simulations shown in Fig. \ref{fig:tree} have been
limited to a total number of tree nodes between 1/10 and 1/3 of the
total number of sites of the support lattice.

While our discussion was focused on electric currents and potentials,
the discovered morphological organization of point-to-point transport
is relevant for other phenomena.  For instance, steady-state diffusion
of particles from a source to a sink is governed by the Laplace
equation whose discretization on a given graph yields a set of linear
equations similar to the Kirchhoff's equations.  In this setting,
electric potentials and currents are substituted by concentrations of
particles and their fluxes, respectively.  Moreover, when the source
concentration is set to $1$, concentration at a given point can
alternatively be interpreted as the splitting probability that a
particle started from that point hits the source point before hitting
the drain point.  In other words, it characterizes the relative
``accessibilities'' of a source and a drain from different points of
the graph \cite{Grebenkov05}.

In summary, the application of a potential difference at two randomly
chosen points of a scale-free network determines a transport structure
composed of two random trees (rooted at these two points) connected
through a quasi-equipotential cluster.  Small currents between the
nodes of this cluster form the plateau region of the density of
currents.  The arborescent character of the resistant part of this
structure implies the multiplicative reduction of the currents in
these trees and thus a power law decay of the density of currents with
the universal exponent $-2$.  Whatever the rules of construction of a
scale-free network (e.g., the choice of $\beta$ and $\gamma$), the
density of currents exhibits the same decay.  Note that similar
structural organizations of the point-to-point transport are found in
various natural and human-engineered complex systems: human vascular
network (with an artery splitting progressively into numerous
capillaries and then merging again to few veins), braided river
networks \cite{Rinaldo2014,Connor2018}, water supply networks and
irrigation systems (with one or several water intake stations
supplying a branching network of pipes and then merging again to one
or several sewage disposal points), to name but a few.  Here, we
showed that such a structural organization is formed spontaneously in
a scale-free network by applying the potential difference to two
arbitrary points.  This observation is expected to help to understand
the morphology, optimality, and robustness of natural point-to-point
transport systems.  An extension of this study to other types of
complex networks such as small-world graphs, presents an interesting
perspective for future research.  More generally, our interpretation
in terms of arborescent structure of links that drive the currents
from source to drain through a large QEP cluster of nodes opens the
door to aggregate conceptualizations of transport processes in complex
networks.



%


%

%

\section*{Methods}


\subsection*{Description of the numerical scheme}
\label{sec:numerics}

Numerical simulations of the scale-free resistor networks are
performed using MATLAB 9.2 (R2017a).  The source (potential set to 1)
and drain (potential set to 0) nodes are randomly picked among nodes
that are separated by a distance of at least 4 nodes.  The system of
linear Kirchhoff's equations \cite{Redner2001}, excluding the source
and drain nodes, reads in a matrix form as $M \times P = S$, where $P$
is the column vector of all $N-2$ unknown potentials, $M$ is the
$(N-2) \times (N-2)$ matrix of coefficients, and $S$ is the column
vector of $N-2$ elements, where each element corresponds to the total
flow exiting the current node $j$ and satisfying the conservation of
mass equation.  The system is solved using the typical ``mldivide''
(or backslash) operator in MATLAB, that computes the solution $P$
using LU decomposition.  Knowing the potential $P$, one can then
compute the current for each link.

\subsection*{Computation of current distribution on a tree}
\label{sec:tree_examples}

We discuss two examples of trees for illustrating the general
arguments of the main text about the distribution of currents.

First, we consider a regular tree with $N$ levels of branching, in
which each node (except the root) is divided into $q$ daughter
branches.  We select the root node as the source and all terminal
nodes as a drain.  The correspondence to the point-to-point transport
can be made by merging this tree to its copy (``reflected tree'') by
connecting pairwise all terminal nodes.  In that case, the root of the
reflected tree is set as the drain.  The distribution of currents is
the same in both cases due to the symmetry.  As this network is
deterministic, the currents are as well deterministic, so that one can
understand the distribution of currents in terms of frequencies of
observation of a given value of the current.  Clearly, such a
distribution is discrete.

Due to the symmetry, currents in all branches of a given level are
identical.  If $\phi_0$ denotes the current in the root branch (of
level $0$), which is single by construction, then the current in any
branch of the level $n$ is just $\phi_0/q^n$, where $q^n$ is the
number of branches at this level.  Then the probability (interpreted
as the frequency of occurrence) for the current $\hat{\phi}$ in a
randomly selected branch is
\begin{equation}
P_n = \P\{ \hat{\phi} = \phi_0/q^n \} = \frac{q^n}{1 + q + \ldots + q^{N-1}} \,,
\end{equation}
from which, denoting $\phi = \phi_0/q_n$, one has
\begin{equation}
\P\{ \hat{\phi} \geq \phi\} = P_0 + P_1 + \ldots + P_n = \frac{q \phi_0/\phi - 1}{q^N - 1} \,.
\end{equation}
Formally treating $\phi$ as a continuous variable, one finds that the
corresponding probability density function bahaves as
\begin{equation}
p(\phi) = - \frac{\partial \P\{ \hat{\phi} \geq \phi\}}{\partial \phi} \propto \frac{1}{\phi^2} \,.
\end{equation}

This computation can be extended a nonregular tree, in which the
number $q_n$ of daughter branches depends on the branching level $n$.
Repeating the above arguments, one gets
\begin{equation}
\P\{ \hat{\phi} \geq \phi\} = A \bigl(1 + q_1 + q_1 q_2 + \ldots + q_1 q_2 \cdots q_n\bigr),
\end{equation}
where $\phi = \phi_0/(q_1 q_2 \cdots q_n)$, and $A = 1 + q_1 + q_1 q_2
+ \ldots + q_1 q_2 \cdots q_{N-1}$ is the normalization constant.
Since all $q_n \geq 2$, one can easily prove that the last term, $q_1
q_2 \cdots q_n$, provides the dominant contribution to this sum.  In
other words, the contribution of the remaining terms is either smaller
or comparable to the last term.  In fact, the following inequality
holds
\begin{equation}
1 + q_1 + q_1 q_2 + \ldots + q_1 q_2 \cdots q_{n-1} \leq q_1 q_2 \cdots q_n .
\end{equation}
To prove it, one can divide both sides by $q_1 q_2 \cdots q_{n-1}$ and
check that
\begin{align*}
& \frac{1}{q_1 q_2 \cdots q_{n-1}} + \ldots + \frac{1}{q_{n-1}} + 1 \leq \frac{1}{2^{n-1}} + \ldots + \frac{1}{2} + 1
 = 2(1 - (1/2)^n) \leq 2 \leq q_n .
\end{align*}
As a consequence, $\P\{ \hat{\phi} \geq \phi\} \propto 1/\phi$, and
thus one gets again the scaling exponent $-2$ for the current density.





\section*{Acknowledgements}

C.N. acknowledges support from the University of Cyprus, through a
starting grant.

\end{document}